\documentclass[preprintnumbers,secnumarabic,onecolumn]{revtex4}
\pdfoutput=1
\usepackage{graphicx}
\usepackage{amsmath,amssymb,mathrsfs,bm}
\usepackage{array}

\newcommand*{\ewgroup}{\ensuremath{SU(2)_L \times U(1)_Y}}
\newcommand*{\trans}{\mathrm{T}}                     
\newcommand*{\unitmatrix}{\bm{1}}
\newcommand*{\tvec}[1]{\ensuremath{\boldsymbol{\mathrm{#1}}}}           
\newcommand*{\tmat}[1]{\underline{#1}}             

\makeatletter       

\renewcommand{\p@subsection}{}

\makeatother

\DeclareMathOperator{\re}{Re}
\DeclareMathOperator{\im}{Im}

\begin{document}

\author{Ernest Ma$^a$}
\author{Markos Maniatis$^b$}

\affiliation{a) Department of Physics and Astronomy, University of California,
Riverside, California 92521, USA}
\affiliation{b) Institut f\"ur Theoretische Physik, University of Heidelberg,
69120 Heidelberg, Germany}

\preprint{UCRHEP-T491, HD-THEP-10-9}

\title{Effective Two Higgs Doublets in Nonminimal Supersymmetric Models}

\begin{abstract}
The Higgs sectors of supersymmetric extensions of the Standard Model have 
two doublets in the minimal version (MSSM), and two doublets plus a singlet 
in two others: with (UMSSM) and without (NMSSM) an extra $U(1)'$.  A very 
concise comparison of these three models is possible if we assume that the 
singlet has a somewhat larger breaking scale compared to the electroweak scale. 
In that case, the UMSSM and the NMSSM become effectively two-Higgs-doublet 
models (THDM), like the MSSM.  As expected, the mass of the lightest CP-even 
neutral Higgs boson has an upper bound in each case.  We find that in the 
NMSSM, this bound exceeds not very much that of the MSSM, unless $\tan(\beta)$ 
is near one.  However, the upper bound in the UMSSM may be substantially 
enhanced.
\end{abstract}

\maketitle

In supersymmetric extensions of the Standard Model, the Higgs sector is 
also extended. In the minimal supersymmetric Standard Model (MSSM; see for 
instance the review~\cite{Martin:1997ns}), there are two Higgs doublets 
$\hat{H}_1$ and $\hat{H}_2$, which are necessary to give masses to both up- and 
down-type fermions and to keep the theory anomaly-free.  For various reasons, 
nonminimal supersymmetric extensions of the Standard Model have also been 
considered. One reason is that in the MSSM superpotential, the term 
$\mu \hat{H}_2^\trans \epsilon \hat{H}_1$ is allowed.  This is called the $\mu$ 
problem, because the value of $\mu$ is {\it \`{a} priori} undetermined and yet 
it has to be adjusted by hand to the electroweak scale, {\em before}
electroweak symmetry breaking occurs.  This coincidence is considered by 
many to be {\em unnatural}.  Another more pragmatic reason is that the mass 
of the lightest CP-even neutral Higgs boson is predicted to be rather small, 
close to the exclusion limits of current collider 
experiments~\cite{Schael:2006cr}.\\

Here we consider two supersymmetric extensions where the $\mu$ term is 
absent, namely the UMSSM~\cite{Fayet:1976et,Fayet:1977yc,Farrar:1978xj,
Kim:1983dt,Cvetic:1995rj,Cvetic:1996mf,Cvetic:1997ky,Keith:1997zb}
and the next-to-minimal supersymmetric Standard Model (NMSSM; see the 
recent reviews~\cite{Maniatis:2009re,Ellwanger:2009dp}).  In the UMSSM, 
an extra $U(1)'$ gauge symmetry is assumed, under which $\hat{H}_{1,2}$ 
transform so that the $\mu$ term is forbidden, but is replaced by 
$f \hat{\chi} (\hat{H}_2^\trans \epsilon \hat{H}_1)$, where $\hat{\chi}$ 
is a Higgs singlet, transforming also under $U(1)'$.  An effective $\mu$ 
term is generated in this model by the spontaneous breaking of $U(1)'$ 
through the vacuum expectation value of $\chi$, i.e. $\mu = f \langle \chi 
\rangle$.  Alternatively in the NMSSM, the $\mu$ term is replaced by 
$f \hat{\chi} (\hat{H}_2^\trans \epsilon \hat{H}_1) + (\kappa/3)\hat{\chi}^3$ 
without any $U(1)'$, but with a $\bm{Z}_3$ symmetry to forbid linear and 
quadratic terms in the superpotential.\\

The MSSM, UMSSM, and NMSSM appear to be quite different in their Higgs sectors.
However, if the singlet in the UMSSM and NMSSM, respectively, gets a large 
vacuum expectation value compared to the electroweak scale, all three models 
become effectively two-Higgs-doublet models at the electroweak scale. 
We can thus compare the three models in terms of their differences in this 
limited sector.  We find that an upper bound exists for the lightest neutral 
scalar Higgs boson in all these models and only in the UMSSM does this upper 
bound exceed that of the MSSM substantially, whereas in the NMSSM, this is 
possible only if $\tan(\beta)$ is near one.  Note that this property of 
the UMSSM was already discussed in~\cite{Keith:1997zb}.  Here we consider 
also the NMSSM.

We make use of the bilinear formalism proposed to describe the general 
THDM~\cite{Nagel:2004sw,Maniatis:2006fs,Nishi:2006tg,Maniatis:2007vn}.  
In this formalism, all gauge-invariant expressions are given in terms 
of four real {\em gauge-invariant bilinears}.  In order to make this article 
self-contained, we review here briefly the usage of the bilinears.  
Consider first the most general potential of two Higgs doublets 
$\Phi_1$,$\Phi_2$ in the conventional notation~\cite{Gunion:1989we}
\begin{equation}
\label{Vconv}
\begin{split}
V =~& 
m_{11}^2 (\Phi_1^\dagger \Phi_1) +
m_{22}^2 (\Phi_2^\dagger \Phi_2) -
m_{12}^2 (\Phi_1^\dagger \Phi_2) -
(m_{12}^2)^* (\Phi_2^\dagger \Phi_1)\\
& +\frac{1}{2} \lambda_1 (\Phi_1^\dagger \Phi_1)^2 
+ \frac{1}{2} \lambda_2 (\Phi_2^\dagger \Phi_2)^2 
+ \lambda_3 (\Phi_1^\dagger \Phi_1)(\Phi_2^\dagger \Phi_2) \\ 
&+ \lambda_4 (\Phi_1^\dagger \Phi_2)(\Phi_2^\dagger \Phi_1)
+ \frac{1}{2} [\lambda_5 (\Phi_1^\dagger \Phi_2)^2 + \lambda_5^* 
(\Phi_2^\dagger \Phi_1)^2] \\ 
&+ [\lambda_6 (\Phi_1^\dagger \Phi_2) + \lambda_6^* 
(\Phi_2^\dagger \Phi_1)] (\Phi_1^\dagger \Phi_1) + [\lambda_7 (\Phi_1^\dagger 
\Phi_2) + \lambda_7^* (\Phi_2^\dagger \Phi_1)] (\Phi_2^\dagger \Phi_2).
\end{split}
\end{equation}
Hermiticity of the Lagrangian requires the parameters 
$m_{12}^2$, $\lambda_{5,6,7}$ to be complex and all other parameters to be real.
Owing to \ewgroup~gauge invariance, only terms of the form 
$(\Phi_i^\dagger \Phi_j)$ with $i,j=1,2$ may occur in the Higgs potential.
The Hermitian, positive semi-definite $2 \times 2$ matrix of all possible 
scalar products of this form may be decomposed in the following 
way~\cite{Maniatis:2006fs,Nishi:2006tg},
\begin{equation}
\label{eq-kmat}
\tmat{K} :=
\begin{pmatrix}
   \Phi_1^{\dagger}\Phi_1 & \Phi_2^{\dagger}\Phi_1 \\
  \Phi_1^{\dagger}\Phi_2 & \Phi_2^{\dagger}\Phi_2
\end{pmatrix}
 = \frac{1}{2} \left( K_0 \unitmatrix_2 + K_i  \sigma_i \right),
\end{equation}
with Pauli matrices $\sigma_i$, $i=1,2,3$, and the convention of summing over 
repeated indices adopted.  Specifically, these real gauge-invariant 
bilinears are defined as
\begin{equation}
\label{eqKphi}
K_0 = \Phi_1^\dagger \Phi_1 + \Phi_2^\dagger \Phi_2, ~~~
K_1 = \Phi_1^\dagger \Phi_2 + \Phi_2^\dagger \Phi_1, ~~~
K_2 = i\Phi_2^\dagger \Phi_1 - i\Phi_1^\dagger \Phi_2, ~~~ 
K_3 = \Phi_1^\dagger \Phi_1 - \Phi_2^\dagger \Phi_2. 
\end{equation}
The matrix $\tmat{K}$ in~\eqref{eq-kmat} is positive semi-definite with 
two conditions for the gauge-invariant bilinears:
\begin{equation}
K_0 \ge 0, \qquad K_\alpha K_\alpha \equiv K_0^2 - K_1^2 - K_2^2 - K_3^2 \ge 0.
\end{equation}
For convenience, we introduce the shorthand vector notation 
$\tvec{K}=(K_1,K_2,K_3)^\trans$.  For any $K_0$ and $\tvec{K}$, it is possible 
to find doublet fields $\Phi_{1,2}$ obeying~\eqref{eqKphi}.  These doublets 
then form an gauge orbit.  In terms of the gauge-invariant bilinears, the 
general THDM potential may be written in the simple form
\begin{equation}
\label{VK}
V = \xi_{\alpha}  K_\alpha + \eta_{\alpha \beta} K_\alpha K_\beta,
\qquad \alpha, \beta = 0,...,3\;,
\end{equation}
where $\xi$ is a real 4-vector and
 $\eta$ is a real and symmetric $4 \times 4$ matrix.
 Expressed in terms of the conventional parameters, these tensors
 read
\begin{equation}
\xi=\frac{1}{2}
\begin{pmatrix}
m_{11}^2+m_{22}^2, & 
- 2 \re(m_{12}^2), &
 2 \im(m_{12}^2), &
 m_{11}^2-m_{22}^2
\end{pmatrix}
\end{equation}
and
\begin{equation}
\label{VK4}
\eta = \frac{1}{4}
\begin{pmatrix}
\frac{1}{2}(\lambda_1 + \lambda_2) + \lambda_3 & 
\re(\lambda_6+\lambda_7) & 
-\im(\lambda_6+\lambda_7) & 
\frac{1}{2}(\lambda_1 - \lambda_2) \\ 
\re(\lambda_6+\lambda_7) & 
\lambda_4 + \re(\lambda_5) & 
-\im(\lambda_5) & \re(\lambda_6-\lambda_7) 
\\ 
-\im(\lambda_6+\lambda_7) & 
-\im(\lambda_5) & \lambda_4 - \re(\lambda_5) & 
-\im(\lambda_6-\lambda_7) \\ 
\frac{1}{2}(\lambda_1 - \lambda_2) & 
\re(\lambda_6-\lambda_7) & 
-\im(\lambda_6 -\lambda_7) & 
\frac{1}{2}(\lambda_1 + \lambda_2) - \lambda_3
\end{pmatrix}.
\end{equation}\\


Let us first recall the MSSM with the Higgs part of
the superpotential
\begin{equation}
W_{\text{MSSM}}^{\text{Higgs}}= \mu (\hat{H}_2^\trans \epsilon \hat{H}_1)
\end{equation}
with Higgs doublet supermultiplets $\hat{H}_1$ and $\hat{H}_2$. 
From the scalar part of the superpotential we get the $F$-terms
and from the gauge interactions we get the $D$-terms, in addition
to the soft supersymmetry breaking terms,
\begin{equation}
\label{MSSMpot}
\begin{split}
V_F^{\text{MSSM}} &= |\mu|^2 (\Phi_1^\dagger \Phi_1+\Phi_2^\dagger \Phi_2) ,\\
V_D^{\text{MSSM}} &= \frac{g_1^2 + g_2^2}{8}(\Phi_1^\dagger \Phi_1- \Phi_2^\dagger 
\Phi_2)^2 +\frac{g_2^2}{2} \big( (\Phi_1^\dagger \Phi_1)(\Phi_2^\dagger \Phi_2)-(\Phi_1^\dagger \Phi_2)
(\Phi_2^\dagger \Phi_1) \big),\\
V_{\text{soft}}^{\text{MSSM}} &= m_{H_1}^2 (\Phi_1^\dagger \Phi_1) + m_{H_2}^2 
(\Phi_2^\dagger \Phi_2)
- (m_3^2 (\Phi_1^\dagger \Phi_2) + H.c.) .
\end{split}
\end{equation}
The Higgs-boson doublets $H_1$ and $H_2$ carry hypercharges $-1/2$ and
$+1/2$, respectively. In~\eqref{MSSMpot} we have changed the convention and 
consider doublets with the same hypercharge,~$+1/2$,
\begin{equation}
\label{hyper}
\Phi_1 = -\epsilon H_1^* ,\quad
\Phi_2=H_2, \quad \text{with }  
\epsilon=
\begin{pmatrix}
0 & 1 \\ -1 & 0
\end{pmatrix} .
\end{equation}
Hence, we
have the potential parameters in conventional 
notation
\begin{gather}
\begin{split}
m_{11}^2 = m_{H_1}^2 + |\mu|^2, \quad
m_{22}^2 = m_{H_2}^2 + |\mu|^2, \quad
m_{12}^2 = m_3^2 ,\\
\lambda_1 =\frac{g_1^2+g_2^2}{4}, \quad
\lambda_2 =\frac{g_1^2+g_2^2}{4} + \delta, \quad
\lambda_3 = \frac{g_2^2 - g_1^2}{4}, \quad
\lambda_4 = - \frac{g_2^2}{2}, \\
\lambda_5=\lambda_6=\lambda_7 = 0.
\end{split}
\end{gather}
Here we have taken into account the generically large 
 $\Phi_2$--$\Phi_2$ self-energy correction from a top--stop loop
which contributes only to $\lambda_2$ in supersymmetric models,
\begin{equation}
\label{epsilon}
\delta= \frac{3 g_2^4 m_t^4}{32 \pi^2 m_W^4} \frac{1}{\sin^4(\beta)}
\ln (1+\frac{m_{\tilde{t}}}{m_t} ) ,
\end{equation}
with $m_t$, $m_W$, $m_{\tilde{t}}$ the masses
of top-quark, W, and stop-quark, respectively.\\

In terms of the bilinears, the Higgs-potential parameters read
\begin{equation}
\xi =
\begin{pmatrix}
\frac{1}{2} (m_{H_1}^2+m_{H_2}^2) + |\mu|^2\\
- \re{m_3^2}\\
\phantom{+} \im{m_3^2}\\
\frac{1}{2} (m_{H_1}^2-m_{H_2}^2)
\end{pmatrix} ,
\quad
\eta = \frac{1}{8}
\begin{pmatrix}
g_2^2+\delta & 0 & 0 & -\delta\\
0 & -g_2^2 & 0 & 0\\
0 & 0 & -g_2^2 & 0\\
-\delta & 0 & 0 & g_1^2 +\delta
\end{pmatrix} ,
\end{equation}
with diagonal quartic coupling matrix at tree level.

The stability condition for 
the MSSM follows from the $D$-flat directions,
i.e. directions in field space with vanishing
$D$-terms in~\eqref{MSSMpot}. Stability 
in these directions is ensured by the quadratic terms, i.e.
\begin{equation}
\xi_0 - \sqrt{\xi_1^2+\xi_2^2} = 
\frac{1}{2} (m_{H_1}^2+m_{H_2}^2) + |\mu|^2 - |m_3^2| > 0 .
\end{equation}

In any THDM Higgs potential, the requirement for the nontrivial
symmetry breaking of $U(1)_Y \times SU(2)_L \to U(1)_{\text{em}}$ 
is reflected by the condition $\xi_0 < | \tvec{\xi}|$~\cite{Maniatis:2006fs}, 
i.e.
\begin{equation}
\label{ewsb}
\frac{1}{2} (m_{H_1}^2+m_{H_2}^2) + |\mu|^2 <
\sqrt{|m_3^2|^2 + \frac{1}{4}(m_{H_1}^2-m_{H_2}^2)^2} .
\end{equation} 
After electroweak symmetry breaking we get the Higgs mass spectrum. 
In the unitary gauge the Higgs-doublet components are
\begin{equation}
\label{unitary}
\Phi_1=
\begin{pmatrix}
0\\ v+ \frac{1}{\sqrt{2}} H_1'
\end{pmatrix},
\qquad
\Phi_2=
\begin{pmatrix}
H^+\\
\frac{1}{\sqrt{2}} (H_2' + i H_3')
\end{pmatrix} ,
\end{equation}
with $v \approx 174$~GeV the vacuum expectation value.  After separating the 
Goldstone mode, we end up with a scalar mass--squared matrix in the basis 
$(H_1', H_2', H_3')^\trans$
\begin{equation}
M^2_{\text{MSSM}}=
\begin{pmatrix}
m_Z^2 c^2_{2 \beta} + \frac{\delta v^2}{4} ( 3 + c_{4\beta} -4 c_{2 \beta} ) &
-\frac{1}{2} m_Z^2 s_{4\beta} + 2 \delta v^2 c_{\beta} s_{\beta}^3 &
0\\
-\frac{1}{2} m_Z^2 s_{4\beta} + 2 \delta v^2 c_{\beta} s_{\beta}^3 &
m_{H^\pm}^2 - m_W^2 + m_Z^2 s^2_{2\beta}+\frac{\delta v^2}{2} s_{2\beta}^2 &
0\\
0 & 0 & 
m_{H^\pm}^2 - m_W^2
\end{pmatrix} .
\end{equation}
Here we employ the abbreviations $c_{\beta}=\cos(\beta), s_\beta=\sin(\beta)$ 
etc. We find the well-known fact 
that the MSSM Higgs potential is CP conserving, 
i.e. we have one neutral CP-odd Higgs boson $A^0$ 
with squared mass $m_{A^0}^2 = m_{H^\pm}^2 - m_W^2$
and two neutral CP-even 
Higgs bosons.
\begin{figure}[t!]
\includegraphics[width=0.6\textwidth]{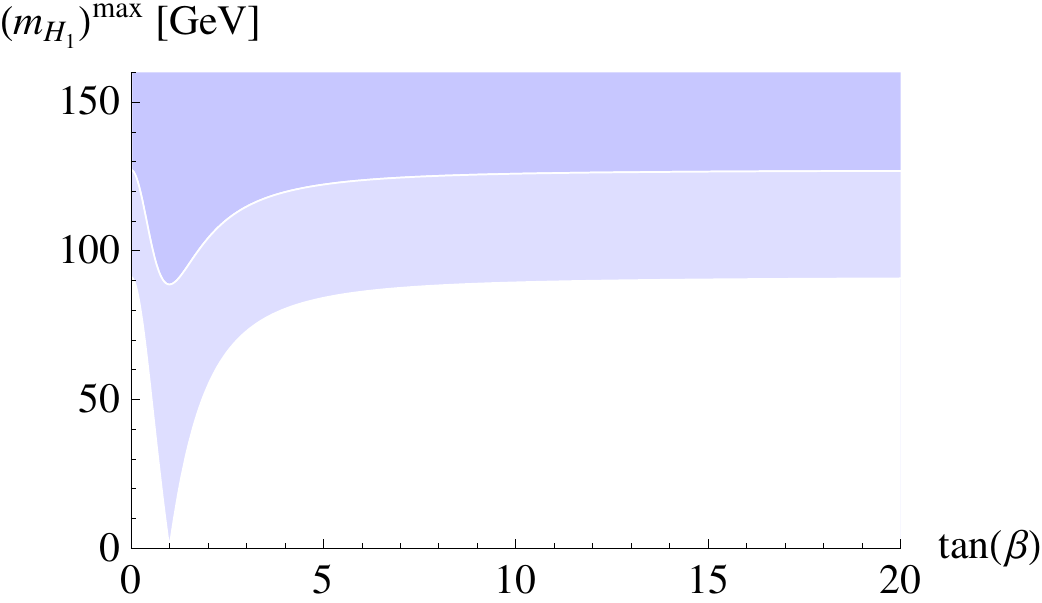}
\caption{\label{mHmaxMSSM} Upper bound on the lightest CP-even Higgs boson 
in the MSSM depending on~$\tan(\beta)$. The light shaded region is excluded 
at tree level and the dark shaded region with the stop loop taken into 
account -- assuming a stop mass of~1~TeV.}
\end{figure}
The upper limit on the lightest CP-even Higgs boson mass squared is
\begin{equation}
m_{H_1}^2 \le m_Z^2 \cos^2(2\beta) + 
\frac{v^2 \delta}{2} \cdot \bigg(1 - 2 \cos(2\beta)+ \cos^2(2\beta) \bigg) .
\end{equation}
In figure~\ref{mHmaxMSSM} we show the upper bound on the lightest neutral 
CP-even Higgs boson depending on~$\tan(\beta)$. The brightly shaded region 
is excluded at tree level, whereas the darkly shaded region is excluded 
when the stop-mass loop is taken into account. The stop mass is set 
to~1~TeV which results in a rather large enhancement of the upper bound.
We confirm from this that even with rather large radiative corrections due 
to a rather large stop mass, we cannot exceed the upper mass bound of 
about~130~GeV in the MSSM.\\


Now let us turn to the nonminimal $U(1)'$ extended supersymmetric model. 
The $U(1)'$ symmetry is violated by the $\mu$ term in the MSSM, but is 
restored by replacing it with 
\begin{equation}
\label{supU}
W_{\text{UMSSM}}^{\text{Higgs}}= f (\hat{H}_2^\trans \epsilon \hat{H}_1) \hat{\chi}
\end{equation}
which is allowed by $U(1)'$ if we assign the respectively charges as given 
in table~\ref{mult}:
\begin{table}[t]
\begin{tabular}{lc|cc|cccc}
\hline
\multicolumn{2}{c|}{supermultiplets} & 
spin-$0$ & spin-$1/2$ & $SU_C(3)$ & $SU_L(2)$ & $U_Y(1)$ & $U(1)'$\\
\hline
& $\hat{H}_1$ & ${H}_1=(H^0_1,\; H^-_1)^\trans$ & 
$\tilde{H}_1=(\tilde{H}^0_1 ,\; \tilde{H}^-_1)^\trans$
& $\mathbf{1}$ & $\mathbf{2}$ & $-1/2$ & $-a$\\
 & $\hat{H}_2$ & ${H}_2=(H^+_u,\; H^0_2)^\trans$ & 
$\tilde{H}_2=(\tilde{H}^+_2 ,\; \tilde{H}^0_2)^\trans$
& $\mathbf{1}$ & $\mathbf{2}$ & $\phantom{+}1/2$ & $a-1$\\
 & $\hat{\chi}$ & $\chi$ & $\tilde{\chi}$ & $\mathbf{1}$ & $\mathbf{1}$ & 
$\phantom{+}0$ & 1\\
\hline
\hline
\end{tabular}
\caption{\label{mult} 
Multiplicities and charges of the Higgs supermultiplets in the $U(1)'$ extended
supersymmetric model.}
\end{table}
The $U(1)'$  gauge coupling is denoted by~$g_x$ and the charges are given in 
terms of the parameter~$a$. The $F$-terms, corresponding to the 
superpotential~\eqref{supU} as well as the $D$- and soft 
supersymmetry-breaking terms read
\begin{equation}
\begin{split}
V_F^{\text{UMSSM}} &=|f|^2 (\Phi_1^\dagger \Phi_2) (\Phi_2^\dagger \Phi_1) +
|f|^2 (\Phi_1^\dagger \Phi_1 + \Phi_2^\dagger \Phi_2) |\chi|^2 ,\\
V_D^{\text{UMSSM}} &=
\frac{g_1^2 + g_2^2}{8}(\Phi_1^\dagger \Phi_1 - \Phi_2^\dagger \Phi_2)^2
+\frac{g_2^2}{2}\big( (\Phi_1^\dagger \Phi_1)(\Phi_2^\dagger \Phi_2)-(\Phi_1^\dagger \Phi_2)(\Phi_2^\dagger \Phi_1) \big)
+\frac{g_x^2}{2} \big(
|\chi|^2 - a \Phi_1^\dagger \Phi_1 - (1-a) \Phi_2^\dagger \Phi_2
\big)^2 ,\\
V_{\text{soft}}^{\text{UMSSM}} &=
m_{H_1}^2 \Phi_1^\dagger \Phi_1 +
m_{H_2}^2 \Phi_2^\dagger \Phi_2 +
m_{\chi}^2 |\chi|^2 -
( f A_f \chi \Phi_1^\dagger \Phi_2 + H.c. ) .
\end{split}
\end{equation}
We again have changed the hypercharge convention such that both doublets
have the same hypercharge~$+1/2$.  Now let us assume that at a high scale 
the $U(1)'$ symmetry is spontaneously broken. With the expansion around the 
vacuum expectation value $u$, $\chi = u + 1/\sqrt{2}(\chi_R + i \chi_I)$, we 
get four-point $\Phi_i^\dagger \Phi_j $--$\chi_R $--$\Phi_k^\dagger \Phi_l $ 
functions, ($i,j,k,l=1,2$) with a $\chi_R$ propagator. For a large singlet 
vacuum expectation value $u$ compared to the electroweak scale, $u >> v$, 
these four-point functions yield effective 
$\Phi_i \Phi_j^\dagger$--$\Phi_k^\dagger \Phi_l$ couplings, similar to the 
effective four-Fermi interaction. The effective four-scalar interactions 
can be immediately calculated from the soft supersymmetry-breaking 
mass term of the singlet with $m_{\chi_R}^2= 2 g_x^2 u^2$.  For the quartic 
couplings in conventional notation, we find in the limit of a large 
vacuum expectation value of the singlet~$u$,
\begin{equation}
\begin{split}
\lambda_1 &= \frac{g_1^2+g_2^2}{4} - \frac{|f| ^4}{g_x^2}
+2 a |f| ^2 ,\\
\lambda_2 &= \frac{g_1^2+g_2^2}{4} - \frac{|f| ^4}{g_x^2}
+2 (1-a) |f| ^2 + \delta ,\\
\lambda_3 &= \frac{g_2^2-g_1^2}{4} + |f| ^2 - \frac{|f| ^4}{g_x^2}  ,\\
\lambda_4 &= -\frac{g_2^2}{2} + |f| ^2, \\
\lambda_5 &= \lambda_6 = \lambda_7 =0.
\end{split}
\end{equation}
From this we get via~\eqref{VK4} immediately the quartic coupling matrix
\begin{equation}
\label{etaUMSSM}
\eta = \frac{1}{4}
\begin{pmatrix}
\frac{g_2^2}{2} + 2|f| ^2 -2 \frac{|f| ^4}{g_x^2}+\frac{\delta}{2} & 0 & 0 & 
(2a-1)|f| ^2-\frac{\delta}{2}\\
0 & -\frac{g_2^2}{2} +|f| ^2 & 0 & 0\\
0 & 0 & -\frac{g_2^2}{2} +|f| ^2 & 0\\
(2a-1)|f| ^2-\frac{\delta}{2} & 0 & 0 & \frac{g_1^2}{2}+\frac{\delta}{2}
\end{pmatrix},
\end{equation}
where we have taken into account the radiative correction~\eqref{epsilon}.

The coupling matrix $\eta$ has the symmetry $\eta_{01}=\eta_{02}=0$, 
$\eta_{12}=\eta_{13}=\eta_{23}=0$, and $\eta_{11}=\eta_{22}$.  
In~\cite{Ma:2009ax} we have already recognized this {\em symbiotic} symmetry, 
which is preserved by the renormalization group equations in the 
nonsupersymmetric case.  The symmetry $\eta_{11}=\eta_{22}$, which originates 
from the $U(1)'$ symmetry, allows us to consider a basis where 
$\xi_2 =0$ without any change of the quartic couplings.

Stability of the Higgs potential is then guaranteed by the quartic terms 
if and only if the parameters fulfill $\eta_{00}+\eta_{33} > 2 |\eta_{03}|$ and
in the case of $\eta_{03}^2 \le (\eta_{11}-\eta_{33})^2$, we also 
require $\eta_{00}+\eta_{11} > \eta_{03}^2/(\eta_{33}-\eta_{11})$.

The matrix of the squared masses follows in the UMSSM as
\begin{equation}
M^2_{\text{UMSSM}} = 
\begin{pmatrix}
m_Z^2 c_{2\beta}^2 + \Delta M_0^{11} & -1/2 m_Z^2 s_{4\beta} + \Delta M_0^{12} & 0\\
-1/2 m_Z^2 s_{4\beta} + \Delta M_0^{12} & m_{H^\pm}^2-m_W^2 + m_Z^2 s_{2\beta}^2 
+ \Delta M_0^{22} & 0\\
0 & 0 & m_{H^\pm}^2-m_W^2 + v^2 |f|^2
\end{pmatrix}
\end{equation}
with
\begin{equation}
\begin{split}
\Delta M_0^{11} &= - \frac{v^2}{4} \big(
8 \frac{|f|^4}{g_x^2} + 2 |f|^2 ( (4-8 a) c_{2\beta} + c_{4\beta} -5 ) 
- \delta \cdot ( 3- 4 c_{2\beta}+c_{4\beta} )
  \big) ,\\
\Delta M_0^{12} &= \frac{v^2}{2} s_{2\beta} \big(
2 |f|^2 (1- 2 a + c_{2\beta})  - \delta \cdot (c_{2\beta}-1 )
\big), \\
\Delta M_0^{22} &= \frac{v^2}{4} \big(
4 |f|^2 c_{2\beta}^2 + \delta \cdot (1 - c_{4\beta} )
\big) .
\end{split}
\end{equation}
\begin{figure}[t!]
\includegraphics[width=0.6\textwidth]{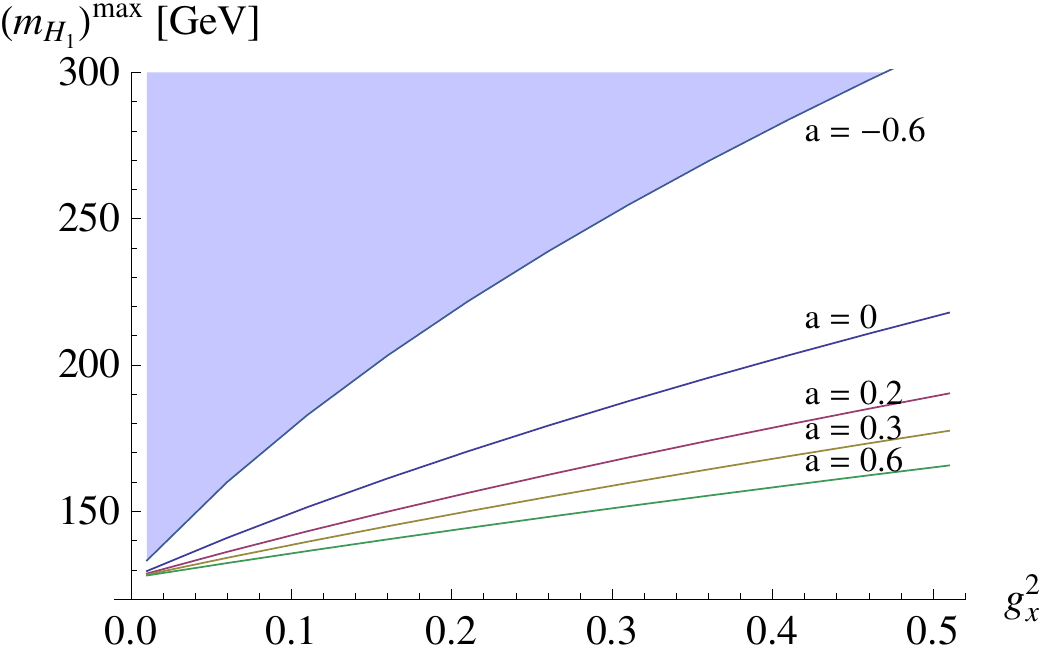}
\caption{\label{mHmaxUMSSM} Upper bound on the lightest CP-even Higgs-boson 
mass in the effective UMSSM depending on the squared coupling $g_x^2$ for 
different values of the charge parameter~$a$. The parameter~$f$ is varied 
in the range $0<f<0.8$ and $0<\beta<\pi/2$ in order to find the maximum 
upper bound.  In the upper shaded region the exclusion region is drawn, 
the border corresponding to a charge~$a=-0.6$.}
\end{figure}
Since the lower right entry does not mix with the other components, we have 
CP conservation in the Higgs sector with a pseudoscalar Higgs mass squared
$m_A^2 = m_{H^\pm}^2-m_W^2 + v^2 |f|^2$.
The upper bound on the lightest of the two CP-even Higgs bosons is
\begin{equation}
m_{H_1}^2 \le 
m_Z^2 c_{2\beta}^2  - \frac{v^2}{4}
  \bigg(
  8 \frac{|f|^4}{g_x^2} 
+ 2 |f|^2 ( (4-8a) c_{2\beta} + c_{4\beta} -5 )
- \delta \cdot (c_{4\beta}-4 c_{2\beta}+3)
  \bigg).
\end{equation}
As pointed out already in~\cite{Keith:1997zb}, the MSSM bound can be exceeded.
The upper bounds of the lightest CP-even Higgs-boson mass are shown
in figure~\ref{mHmaxUMSSM} depending on the squared coupling~$g_x^2$ for
different values of the charge parameter~$a$.  In this plot the upper bound 
is obtained by varying the parameter~$f$ in the range $0<f<0.8$ and 
$0<\beta<\pi/2$. We see that the upper bound on the lightest CP-even Higgs 
boson may substantially exceed the bound of the MSSM.\\


Next we want to study the NMSSM with a hierarchy of breaking scales.
In the NMSSM the $\mu$-term of the MSSM is replaced by
\begin{equation}
W_{\text{NMSSM}}^{\text{Higgs}} =
f (\hat{H}_2^\trans \epsilon \hat{H}_1) \hat{\chi} + \frac{\kappa}{3} \hat{\chi}^3
\end{equation}
In this way the extra $U(1)'$ symmetry is explicitly broken 
and there remains only a~$\bm{Z}_3$-symmetry. Note that
both complex parameters~$f$ and $\kappa$ are dimensionless.
The $F$-terms read
\begin{equation}
\label{potNMSSM}
\begin{split}
V_F^{\text{NMSSM}} &= 
|f| ^2 |\chi|^2 \left( (\Phi_1^\dagger \Phi_1) + (\Phi_2^\dagger \Phi_2) \right) 
+ \left|\kappa \chi^2 - f (\Phi_1^\dagger \Phi_2) \right|^2\,, \\
V_{\text{soft}}^{\text{NMSSM}} &=
m_{H_1}^2 \Phi_1^\dagger \Phi_1 +
m_{H_2}^2 \Phi_2^\dagger \Phi_2 +
m_{\chi}^2 |\chi|^2 -
( f A_f \chi \Phi_1^\dagger \Phi_2  + \frac{\kappa}{3} A_\kappa \chi^3 + H.c. ) .
\end{split}
\end{equation}
with unchanged $D$-terms compared to the MSSM. Note that we have changed the 
hypercharge convention via~\eqref{hyper}.  Expanding the singlet $\chi$ 
around its vacuum expectation value~$u$, we get effective 4-scalar interactions.
From the soft term for the singlet we get the mass squared of its real 
component $m_{\chi_R}^2 = 4 |\kappa|^2 u^2$ in this case.  The parameters of the 
effective two-Higgs potential have in conventional notation the form
\begin{equation}
\begin{split}
\lambda_1 &= \frac{g_1^2+g_2^2}{4} - \frac{|f| ^4}{2 |\kappa|^2} ,\\
\lambda_2 &= \lambda_1 + \delta,\\
\lambda_3 &= \frac{g_2^2-g_1^2}{4} - \frac{|f| ^4}{2 |\kappa|^2} ,\\
\lambda_4 &= -\frac{g_2^2}{2} + \frac{|f| ^2}{2} ,\\
\lambda_5 &= - \frac{f^2 \kappa^*}{2 \kappa} ,\\
\lambda_6 &= \lambda_7 = \frac{1}{2}\frac{f}{\kappa}|f|^2 ,
\end{split}
\end{equation}
where we drop all contributions which vanish in the limit of
a large vacuum-expectation-value of the singlet~$u$. In terms of the 
bilinears, the effective THDM Higgs potential has the parameters
\begin{equation}
\label{etaNMSSM}
\eta = \frac{1}{4}
\begin{pmatrix}
\frac{g_2^2}{2} - \frac{|f| ^4}{|\kappa|^2}+\frac{\delta}{2} &
\frac{|f| ^2}{|\kappa|^2} \re(f \kappa^*) & 
- \frac{|f| ^2}{|\kappa|^2} \im(f \kappa^*) & 
-\frac{\delta}{2}\\
\frac{|f| ^2}{|\kappa|^2} \re(f \kappa^*) & 
-\frac{g_2^2}{2} + \frac{|f| ^2}{2} -\frac{1}{2} \re(f^2 \kappa^*/\kappa) &
\frac{1}{2} \im(f^2 \kappa^* / \kappa) &
0 \\
- \frac{|f| ^2}{|\kappa|^2} \im(f \kappa^*) & 
\frac{1}{2} \im(f^2 \kappa^* / \kappa) &
-\frac{g_2^2}{2} + \frac{|f| ^2}{2} + \frac{1}{2}\re(f^2 \kappa^*/\kappa) &
0\\
-\frac{\delta}{2} &
0 &
0 &
\frac{g_1^2}{2} + \frac{\delta}{2}
\end{pmatrix}
\end{equation}

\begin{figure}[t!]
\includegraphics[width=0.6\textwidth]{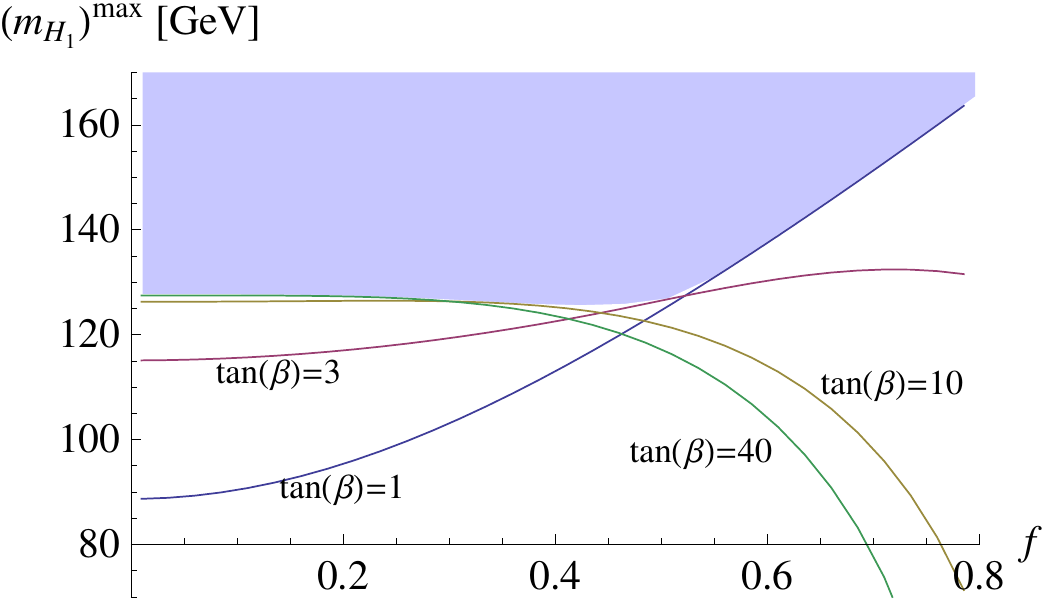}
\caption{\label{mHmaxNMSSM} Upper bound on the lightest CP-even Higgs-boson 
mass, depending on the parameter $f$ in the effective THDM originating from
the NMSSM. In this plot $\kappa$ is varied in the range $0<\kappa<0.8$ and 
$\tan(\beta)$ fixed to different values. In the upper shaded region the 
exclusion region is drawn for arbitrary values of $\tan(\beta)$ in the 
allowed range.}
\end{figure}

For the scalar mass squared matrix we get for the NMSSM in the basis 
$(H_1', H_2', H_3')$ (see~\eqref{unitary})
\begin{equation}
M^2_{\text{NMSSM}} = 
\begin{pmatrix}
m_Z^2 c_{2\beta}^2 + \Delta M_0^{11} & -1/2 m_Z^2 s_{4\beta} + \Delta M_0^{12} & 
\Delta M_0^{13}\\
-1/2 m_Z^2 s_{4\beta} + \Delta M_0^{12} & m_{H^\pm}^2 -m_W^2 + m_Z^2 s_{2\beta}^2 
+ \Delta M_0^{22} & \Delta M_0^{23}\\
\Delta M_0^{13} & \Delta M_0^{23} & m_{H^\pm}^2 -m_W^2 + \Delta M_0^{33}
\end{pmatrix}
\end{equation}
with
\begin{equation}
\begin{split}
\Delta M_0^{11} &= \frac{v^2}{4 |\kappa|^2} \bigg(-4 |f|^4 + 8 |f|^2 
\re(f \kappa^*) s_{2\beta}
+4  \im^2(f \kappa^*) s_{2\beta}^2 + \delta |\kappa|^2 (3 + c_{4\beta} 
-4 c_{2\beta})\bigg) ,\\
\Delta M_0^{12} &= \frac{v^2}{4} \bigg(4 \frac{|f|^2}{|\kappa|^2}  
\re(f \kappa^*)c_{2\beta} + |f|^2 - \re(f^2 \kappa^*/\kappa) s_{4\beta} 
+\delta (2 s_{2\beta}  - 1)\bigg), \\
\Delta M_0^{13} &= \frac{v^2}{2} \bigg( \im(f^2 \kappa^*/\kappa) s_{2\beta} -
2 \frac{|f|^2}{|\kappa|^2} \im(f \kappa^*)\bigg), \\
\Delta M_0^{22} &= \frac{v^2}{4} \bigg( |f|^2 (1+c_{4\beta}) -2 c_{2\beta}^2 
\re(f^2 \kappa^*/\kappa) + \delta (1 - c_{4\beta})
\bigg),\\
\Delta M_0^{23} &= \frac{v^2}{2} \bigg(\im(f^2 \kappa^*/\kappa) 
c_{2\beta} \bigg), \\
\Delta M_0^{33} &=\frac{v^2}{2} \bigg( |f|^2 +\re(f^2 \kappa^*/\kappa) \bigg).
\end{split}
\end{equation}
Obviously the three scalar Higgs bosons mix, i.e. we do not have CP 
conservation in general.  In the special case of vanishing imaginary parts 
of $f \kappa^*$ and $f^2 \kappa^*/\kappa$, for instance with real values of 
$\lambda$ and $\kappa$, the mass squared matrix becomes block diagonal and 
we have CP conservation.  The pseudoscalar squared mass is then
\begin{equation}
m_A^2 = m_{H^\pm}^2 -m_W^2 + \Delta M_0^{33}.
\end{equation}
The lightest CP-even Higgs scalar $H_1$
has in this case an upper bound,
\begin{equation}
m_{H_1}^2 \le 2 v^2 \bigg(2 \eta_{00} + \eta_{11} + \eta_{33} + 4 \eta_{01} 
\sin(2\beta) + 4 \eta_{03} \cos(2\beta) +(\eta_{33}-\eta_{11}) \cos(4\beta) 
\bigg) .
\end{equation} 
If we plug in the quartic couplings in the NMSSM from~\eqref{etaNMSSM}, we
get the explicit upper bounds as shown in~Fig.~\ref{mHmaxNMSSM} depending
on~$f$ for different values of~$\tan(\beta)$.   In this figure for given~$f$ 
we vary $\kappa$ in the range $0<\kappa<0.8$ such as to find the maximum 
upper bound. The shaded region is excluded for additional variation 
of~$\beta$ in the range $0<\beta<\pi/2$, which corresponds to
a positive value for the Higgs-doublet vacuum expectation value. 
Note that the NMSSM Higgs potential is always bounded from below for 
nonvanishing parameters $f$ and $\kappa$.

We see that a large CP-even Higgs boson mass exceeding the MSSM bound is 
possible only for large values of the parameter~$f$ and $\tan(\beta)$ near 
one. On the other hand it is well-known from the renormalization group 
equations that there is a Landau pole for large values 
of~$f$~\cite{Miller:2003ay}.\\

To summarize, we have considered the Higgs sectors of three favored 
supersymmetric extensions of the Standard Model, namely the MSSM, the UMSSM, 
and the NMSSM. We find that in the limit of a hierarchy of breaking scales, 
with the singlet vacuum expectation value scale~$u$ somewhat larger than 
the electroweak breaking scale~$v \approx 174$~GeV,  the lightest CP-even 
Higgs-boson mass is still bounded.  In the NMSSM, this upper 
bound does not exceed much that of the MSSM, unless $\tan(\beta)$ is near one, 
whereas in the UMSSM it is softened substantially in general for moderate 
values of the gauge coupling~$g_x$ and the charge parameter~$a$.\\

Let us also mention the very recent approach~\cite{Delgado:2010uj}, 
where the $\mu$ term is kept in addition to singlet terms
in the superpotential,
$W_{\text{S-MSSM}}^{\text{Higgs}} = (\mu + f \hat{\chi})(\hat{H}_2^\trans \epsilon 
\hat{H}_1) + \frac{\mu_S}{2} \hat{\chi}^2$.
In this way the upper bound of the lightest CP-even Higgs boson is shown to 
exceed that of the MSSM.  In the discussions of the NMSSM 
in~\cite{Tobe:2002zj,Birkedal:2004zx,Franceschini:2010qz}, 
it is shown that allowing the superpotential parameter~$f$ to be 
arbitrarily large up to the GUT scale could result in a large enhancement 
of the CP-even Higgs-boson mass -- a result which we confirm for the case 
of a breaking hierarchy; see Fig.~\ref{mHmaxNMSSM}.

\acknowledgments{M.M. thanks the UC Riverside for the great hospitality 
during his recent visit. The work of E.M. was supported in part by the 
U.~S.~Department of Energy under Grant No.~DE-FG03-94ER40837.  The work 
of M.M. was funded by Deutsche Forschungsgemeinschaft, project number NA296/5-1.
}

%
%

\bibliographystyle{unsrt}

\end{document}